\documentclass[3p]{elsarticle}
\usepackage{amsfonts}
\usepackage{amsmath}
\usepackage{amssymb}
\usepackage{graphicx}%

\biboptions{numbers,sort&compress}

\makeatletter
\def\ps@pprintTitle{%
 \let\@oddhead\@empty
 \let\@evenhead\@empty
 \def\@oddfoot{}%
 \let\@evenfoot\@oddfoot}
\makeatother

\begin{document}

\title{Chirality in a quaternionic representation of the genetic code}

\author[ifly,utn]{C. Manuel Carlevaro}
\ead{manuel@iflysib.unlp.edu.ar}
\author[ifly,jaur]{Ramiro M. Irastorza}
\ead{rirastorza@iflysib.unlp.edu.ar}
\author[ifly,gamefi]{Fernando Vericat\corref{cor}}
\ead{vericat@iflysib.unlp.edu.ar}

\cortext[cor]{Corresponding author}
\address[ifly]{Instituto de F\'isica de L\'iquidos y Sistemas Biol\'ogicos, 59 Nro. 780, 1900, La Plata, Argentina.}
\address[utn]{Universidad Tecnol\'ogica Nacional, Facultad Regional Buenos Aires, Mozart Nro. 2300, C14071VT, Buenos Aires, Argentina.}
\address[jaur]{Instituto de Ingenier\'ia y Agronom\'ia, Universidad Nacional Arturo Jauretche, 1888 Florencio Varela, Buenos Aires, Argentina.}
\address[gamefi]{Grupo de Aplicaciones Matem\'aticas y Estad\'isticas de la Facultad de Ingenier\'ia (GAMEFI), Universidad Nacional de La Plata, Calle 115 y 48, 1900 La Plata, Argentina.}
 
\begin{abstract}
A quaternionic representation of the genetic code, previously reported by the
authors, is updated in order to incorporate chirality of nucleotide bases and
amino acids. The original representation assigns to each nucleotide base a
prime integer quaternion of norm 7 and involves a function that associates
with each codon, represented by three of these quaternions, another integer
quaternion (amino acid type quaternion) in such a way that the essentials of
the standard genetic code (particulaty its degeneration) are preserved. To
show the advantages of such a quaternionic representation we have, in turn,
associated with each amino acid of a given protein, besides of the type
quaternion, another real one according to its order along the protein (order
quaternion) and have designed an algorithm to go from the primary to the
tertiary structure of the protein by using type and order quaternions. In this
context, we incorporate chirality in our representation by observing that the
set of eight integer quaternions of norm 7 can be partitioned into a pair of
subsets of cardinality four each with their elements mutually conjugates and
by putting they in correspondence one to one with the two sets of enantiomers
(\textit{D} and \textit{L}) of the four nucleotide bases adenine, cytosine,
guanine and uracil, respectively. Thus, guided by two diagrams -specifically
proposed to describe the hypothetical evolution of the genetic codes
corresponding to both of the chiral systems of affinities: \textit{D}%
-nucleotide bases/L-amino acids and \textit{L}-nucleotide bases/D-amino acids
at reading frames $5%
\acute{}%
\rightarrow3%
\acute{}%
$ and $3%
\acute{}%
\rightarrow5%
\acute{}%
$, respectively- we define functions that in each case assign a \ L- (D-)
amino acid type integer quaternion to the triplets of \textit{D-}
(\textit{L-}) bases. The assignation is such that for a given\ D-amino acid,
the associated integer quaternion is the conjugate of that one corresponding
to the enantiomer L. The chiral type quaternions obtained for the amino acids
are used, together with a common set of order quaternions, to describe the
folding of the two classes, L and D, of homochiral proteins.\bigskip\bigskip


\end{abstract}
\maketitle

\section{Introduction}

Homochirality of nucleic acids and proteins is one of the attributes that
characterize life on the Earth\cite{Palyi1,Palyi2}. At present time all
the living organisms in our planet have nucleic acids (DNA, RNA, etc.) with
nucleotide bases that just take, between their two possible chiral forms, the
one usually labelled as \textit{D} (for dextro or right-handed); also, their
proteins are chains of amino acids all of which are enantiomers of
(exclusively) the class L (for levo, or left-handed) except by the amino acid
glycine that is not a chiral molecule. To understand why the combination
\textit{D} for nucleotide bases and L for amino acids (and not any other)
occurs in living systems is one of the greatest quests of Biology. The
attempts to answer this question involves either biotic or abiotic arguments.
In general the hypotheses of the first type assume that homochirality is
determined by biological necessity and that it is the result of diverse
selection mechanisms. Among the abiotic hypotheses we have those that propose
a "frozen accident" as the responsible of the homogeneous chirality and those
that assume the existence of some asymmetric force that selects just one of
the chiral forms\cite{Bonner1,Jorissen1,Goodman1}. The problem with these
theories is that, in general, they can not be experimentally checked. This
difficulty is (at least partially) overcome by some hypotheses that claim that
homochirality and the universal genetic code arose closely related. More
precisely, that the genetic code, its translation direction \ and
homochirality emerged through a common natural process of
selection\cite{Root-Bernstein1}. This approach has the advantage that it can
incorporate the idea of an ancestral direct affinity between amino acids and
nucleotide triplets which, these last, further acquire new functions, within a
more modern translation machinery, in the form of codons and anticodons. The
goal is that this early affinity between triplets of nucleotide bases and
amino acids can at present be studied in standard laboratories by simply
synthesizing small RNA-oligonucleotides\cite{Yarus1}. This way preference of
L-amino acids by \textit{D}-bases triplets has been demonstrated\cite{Yarus2,Yarus3,Yarus4,Yarus5,Yarus6,Hobish1,Saxinger1,Walker1,Root-Bernstein2}. Moreover, the non-biological affinity of D-amino
acids by \textit{L}-codons is also observed\cite{Yarus2,Root-Bernstein2,Profy1}. Besides, the hypothesis of coevolution
of homochirality with the genetic code, makes a number of additional
predictions such as the possibility that each codon can encode at least two
different amino acids (according with the reading direction: $5\acute{}\rightarrow 3 \acute{}$ or $3 \acute{} \rightarrow 5 \acute{}$) so the eventual existence of more than a single code becomes
possible\cite{Root-Bernstein2}. We must mention, however, that, despite the
possibility of existence of ancestral \textit{L}-ribonucleic sites for D-amino
acids, there is also experimental evidence of the impossibility of
incorporating D-amino acids in protein structures in present biosynthetic
pathways. Specifically it has been reported chiral discrimination during the
aminoacylation in the active site of the aminoacyl tRNA synthetase and also
during the peptide bond formation\ in the ribosomal peptidyl transferase
center\cite{Nandi1,Englander1}.

The aim of this article is to show how a mathematical representation of the
genetic code recently reported by us\cite{Carlevaro1} can naturally
incorporate chirality in such a way that the resulting description be
consistent with many of the previous observations. Our original representation
is guided by a diagram that we have proposed to sketch the evolution of the
genetic code (see Figure 2 in next Section). The diagram is based on
pioneering ideas by Crick\cite{Crick1,Crick2} and includes the physical
concept of broken symmetry\cite{Hornos1,Maddox1,Stewart1} in a very simple form
that resembles the energy levels of an atom. The representation uses Hamilton
quaternions\cite{Hamilton1,Hamilton2} as main tool. These mathematical
objects are a sort of generalization of the complex numbers and obey an
algebra in many aspects similar to theirs but with the very important (for our
purposes) property that the product is, in general, non commutative. In
addition, the quaternions are ideal for representing spatial
rotations\cite{Altmann1} with important advantages over the classical matrix representation.

In our quaternionic representation we assign to each amino acid in a given
protein two quaternions: an integer one according to which one of the 20
standard amino acids it is (type quaternion) and a real one that determines
its order inside the protein primary structure (order quaternion). The type
quaternions are obtained as a quaternionic function of the codons where each
of the three nucleotides bases is associate to an integer quaternion. The form
of this function being inspired by the code evolution diagram. The order
quaternions, on the other hand, play a fundamental role in relation with the
folding of the protein\cite{Creighton1,BenNaim1}.

The integer quaternions that we choose to associate with the nucleotides bases
belong to a maximum cardinality subset of the set%

\[
\mathbf{H}_{7}\left(
\mathbb{Z}
\right)  =\left\{  \left(  a_{0},a_{1},a_{2},a_{3}\right)  :a_{0},a_{1}%
,a_{2},a_{3}\in%
\mathbb{Z}
\text{; }a_{0}^{2}+a_{1}^{2}+a_{2}^{2}+a_{3}^{2}=7\text{, }a_{0}>0\text{
and\ even}\right\}
\]
with the property that it does not contain pairs of conjugate quaternions. The
set $\mathbf{H}_{7}\left(
\mathbb{Z}
\right)  $ has $7+1=8$ elements\cite{Davidoff1} and so the chosen subset has
$4$ quaternions as it should be. This suggest to partition the set
$\mathbf{H}_{7}\left(
\mathbb{Z}
\right)  $ in the form $\mathbf{H}_{7}\left(
\mathbb{Z}
\right)  =\mathbf{H}_{7;D}\left(
\mathbb{Z}
\right)  \bigcup\mathbf{H}_{7;L}\left(
\mathbb{Z}
\right)  $ (where the four quaternions of $\mathbf{H}_{7;D}\left(
\mathbb{Z}
\right)  \ $and the four quaternions of $\mathbf{H}_{7;L}\left(
\mathbb{Z}
\right)  $ are mutually conjugates) and to associate the elements of
$\mathbf{H}_{7;D}\left(
\mathbb{Z}
\right)  \ $and $\mathbf{H}_{7;L}\left(
\mathbb{Z}
\right)  $ with the four D-nucleotide bases and the four L-nucleotide bases of
D-RNA and L-RNA molecules, respectively. This way, chirality can be included
in our formalism.

In the next Section two diagrams for the evolution of the genetic code are
presented assuming that the two possible combinations ($D$, L) and ($L$, D)
for the chirality of bases triplets and amino acid were present since a
beginning. The diagrams frozen at certain step displaying two different
genetic codes; the one that corresponds to the combination ($D$, L) giving the
present day living systems standard code. Guided by these diagrams, in Section
III, we consider the quaternionic representation of both codes and assign type
quaternions to L- and D-amino acids in such a way that, for a given\ D-amino
acid, the associated integer quaternion is the conjugate of the corresponding
to the enantiomer L. Section IV is devoted to show the advantages of the
quaternionic representation by considering the folding of L- and D-proteins.
Some remarks are finally made in Section V.

\section{Chiral diagrams for the evolution of the genetic code}

Here we generalize, by including chirality, the diagram proposed in Ref.
\cite{Carlevaro1} to describe the genetic code evolution. First we take into
account that Miller-Urey like experiments on the synthesis of organic
molecules in a primordial environment\cite{Miller1,Parker1} show the
formation of amino acids in racemic mixtures. Also, as we have already
mentioned, by studying the affinity of amino acids with small
RNA-oligonucleotides the preference of \textit{D}- \ and L-bases triplets by
L- and D-amino acids, respectively, has been observed\cite{Root-Bernstein2},
so we assume that from the beginnings, the two chirality systems
(\textit{D}-bases/L-amino acids) and (\textit{L}-bases/D-amino acids) evolve
independent one of the other. Moreover, as we describe next, the proposed
evolution diagrams for the two systems are very similar since we assume that
the changes in the corresponding genetic codes are basically independent of
the chirality. Chirality just manifests in settling the two affinity systems.

\begin{figure}[ht]
 \centering
 \includegraphics[scale=0.70]{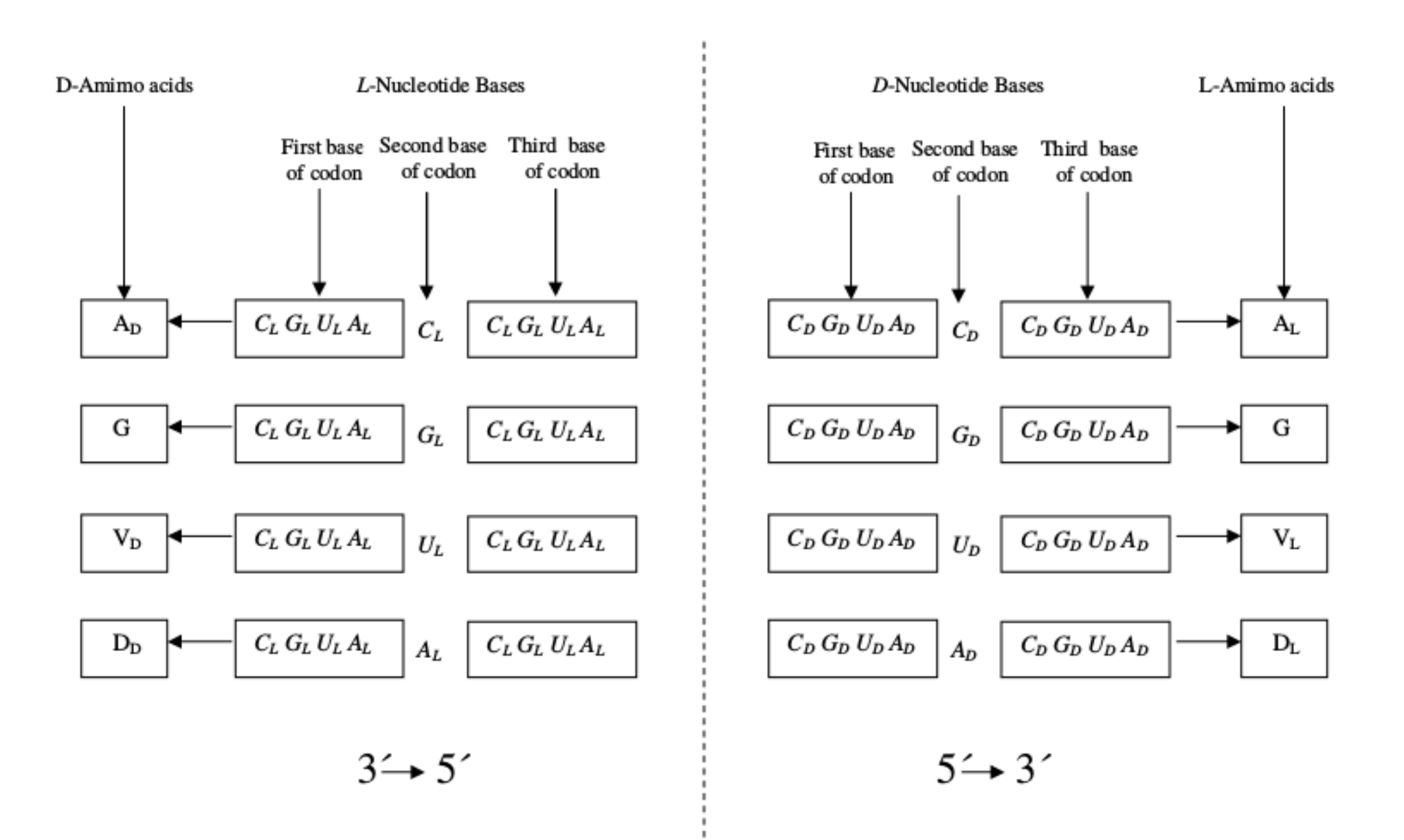}
 \caption{Diagram for the beginnings of the genetic code evolution.
The one letter convention for amino acids is used. The two chiral combinations
(\textit{D}-bases/L-amino acids and \textit{L}-bases/D-amino acids) are
identified with adequate subindices (see text). Note the achirality of the
amino acid glycine (G). The translational direction considered in each case is
also shown. Rectangles with four bases imply fourfold degeneration with
respect to those ones, so at that moment each of the four considered amino
acids was encoded by $16$ triplets.}
\label{fig-01}
\end{figure}

To construct our diagram of the code evolution in Ref. \cite{Carlevaro1} we
have followed Crick\cite{Crick2} and have considered that, in the first
evolution steps, only the second base of the bases triplets was effective in
codifying (binding) amino acids. Here we assume that this fact is valid for
the \textit{L}-bases triplets as well as for the \textit{D}-ones. 

\begin{figure}[th!]
 \centering
 \includegraphics[scale=0.65]{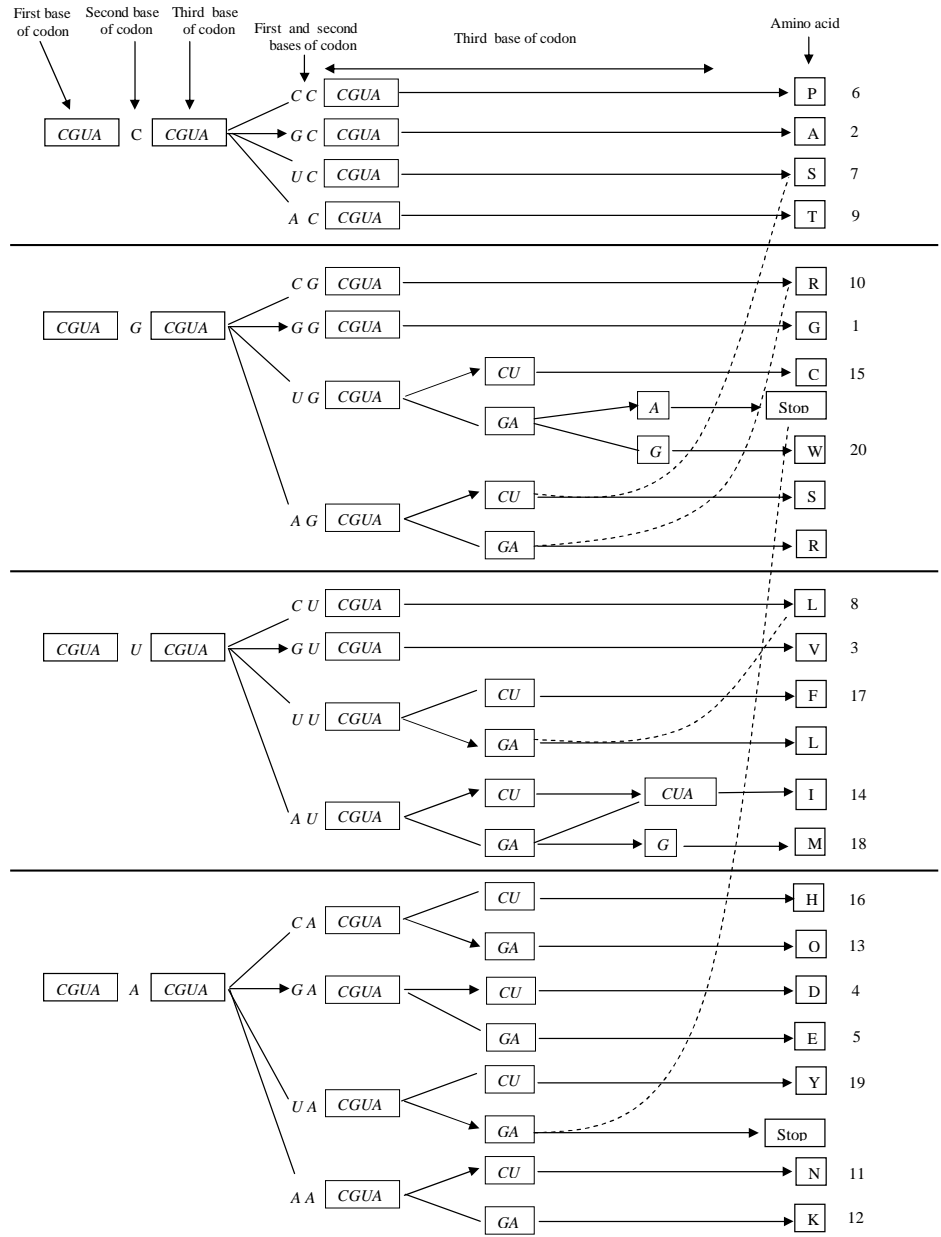}
 \caption{Complete diagram of the genetic code evolution for the
chiral combination \textit{D}-bases/L-amino acids. The one letter convention
for amino acids is used and, for simplicity, the chirality subindices in bases
and amino acids have been dropped. The direction of the temporal evolution is
from left to right. Rectangles with two or more bases implies degeneration
with respect to those ones. The broken lines link different sets of codons
that encode the same amino acid in the case of sixfold degeneration. Arrows
and common lines indicate what codons follow codifying the same amino acid and
what will start to codify a new one, respectively, after the symmetry is
broken (see text). The natural numbers at the right of the amino acids give
the temporal order of the amino acids in the Trifonov consensus
scale\cite{Trifonov1}.}
\label{fig-02}
\end{figure}

\begin{figure}[h!]
 \centering
 \includegraphics[scale=0.63]{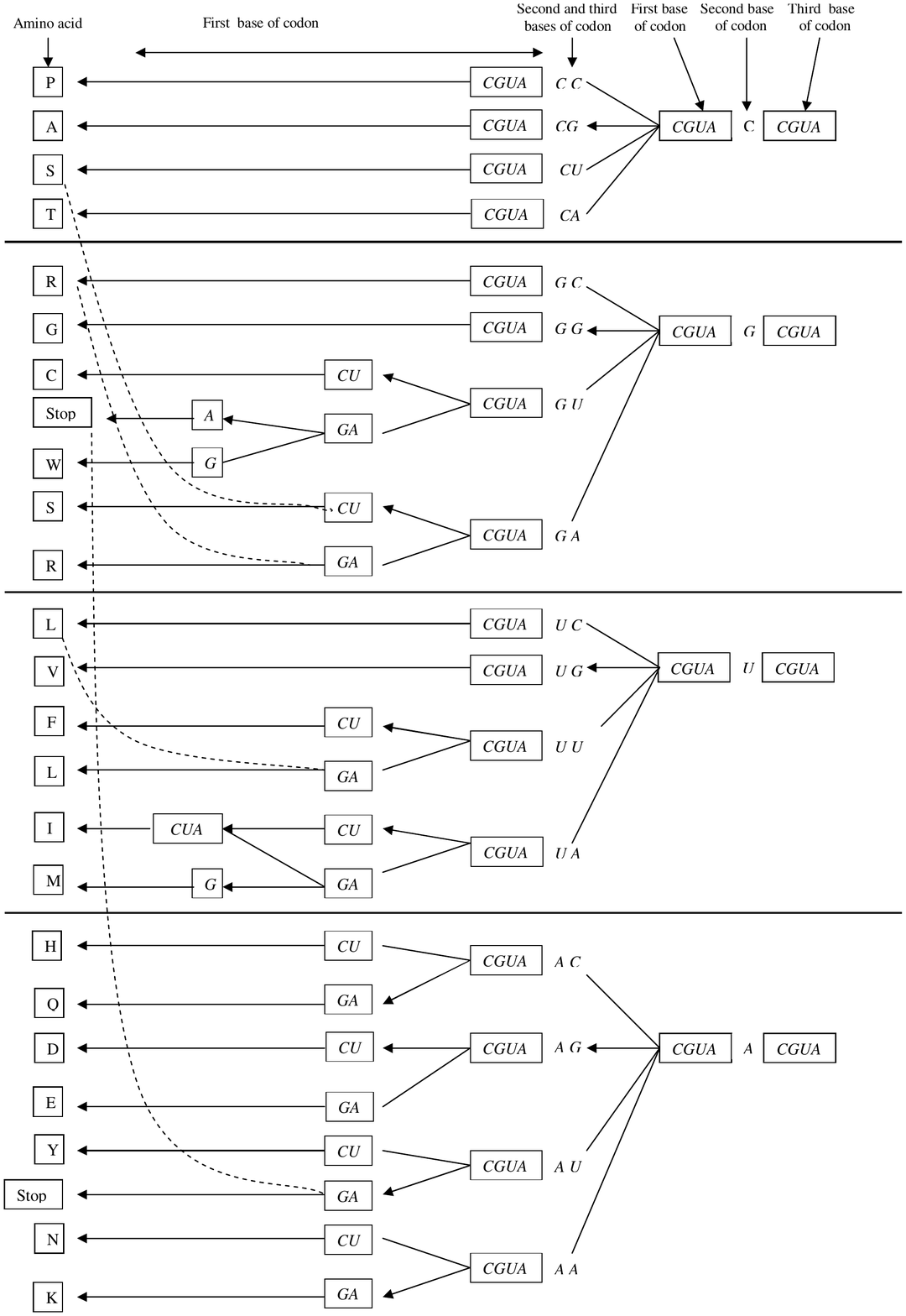}
 \caption{Complete diagram of the genetic code evolution for the
chiral combination \textit{L}-bases/D-amino acids. The one letter convention
for amino acids is used and, for simplicity, the chirality subindices in bases
and amino acids have been dropped. The direction of the temporal evolution is
from right to left. Rectangles with two or more bases implies degeneration
with respect to those ones. The broken lines link different sets of codons
that encode the same amino acid in the case of sixfold degeneration. Arrows
and common lines indicate what codons follow codifying the same amino acid and
what will start to codify a new one, respectively, after the symmetry is
broken (see text).}
\label{fig-03}
\end{figure}
\newpage

Accordingly only four L- (D-) amino acids could be codified, each one by one of the four
enantiomers $C_{\text{\textit{I}}}$ (\textit{I}-cytosine),
$G_{\text{\textit{I}}}$ (\textit{I}-guanine), $U_{\text{\textit{I}}}$
(\textit{I}-uracil)\ and $A_{\text{\textit{I}}}$ (\textit{I}-adenine)
(\textit{I}$=D$ ($L$), respectively) independently of which the first and
third bases are. In Figure \ref{fig-01}, that sketches this first step of the codes
evolution, this fact is denoted with a rectangle containing the four letters.
This is consistent with Crick\'{}s suggestion that only a few amino acids were coded at the beginning.

According with the diagram, $C_{\text{\textit{I}}}$ would codify J-alanine
(A$_{\text{J}}$); $G_{\text{\textit{I}}}$, glycine (G); $U_{\text{\textit{I}}%
}$, J-valine (V$_{\text{J}}$) and $A_{\text{\textit{I}}}$ J-aspartic acid
(D$_{\text{J}}$) (\textit{I},J = \textit{D},L or \textit{L},D) whatever the
first and third bases are. \ It is worth noting here that the four amino acids
that we assume were the first ones to be codified are the first four in the
Trifonov\cite{Trifonov1} consensus temporal order scale for the appearance of
the amino acids (column of natural numbers in Figure 2). The four amino acids
A, G, V and D were also the first four that appeared under simulation of the
primitive earth conditions in Miller experiments\cite{Miller1,Parker1}.
We must also point out\ the two reading frames we are considering: $5 \acute{} \rightarrow 3 \acute{}$ for \textit{D}-triplets and $3\acute{}\rightarrow 5 \acute{}$ for \textit{L}-triplets.

In figures \ref{fig-02} and 3 we show how the evolution follows for the pairs
(\textit{D},L) or (\textit{L},D), respectively. Since confusion is not
possible, we have ignored the subindices that denote chiral class for notation simplicity.

As the left (right) part of diagram of Figure 2 (3) shows, our version of the
primitive code is highly degenerate: in principle each of the four amino
acids, A$_{\text{J}}$,G,V$_{\text{J}}$ and D$_{\text{J}}$, could be encoded by
$4^{2}=16$ codons (see also Figure 1). Physically the idea of degeneration is
closely related with the concept of symmetry and a very illustrative form to
think about these concepts is by doing an analogy with the energy levels of an
atom. In our case we would have four levels indexed each one with the letter
corresponding to the second codon base, say $C_{\text{\textit{I}}}$,
$G_{\text{\textit{I}}}$, $U_{\text{\textit{I}}}$ and $A_{\text{\textit{I}}}$
(main quantum number). We thus assume that, as the code evolves, the symmetry
that causes that the amino acid codification be independent of the first
(third) base of the codon, disappears. Because of this symmetry breaking, a
part of the degeneration also disappears. In the diagrams each of the four
initial levels splits into four new levels, one for each of the possible bases
($C_{\text{\textit{I}}}$, $G_{\text{\textit{I}}}$, $U_{\text{\textit{I}}}$ and
$A_{\text{\textit{I}}}$) at the first (third) place of the codon (secondary
quantum number). Now we have a total of $16$ levels indexed each one by two
letters (the first and second (second and third) bases of the codon). Each
level is fourfold degenerate in the codons third (first) base. One of the new
levels follows codifying the same amino acid as before that the level splits
whereas the other three codify a new amino acid each. We indicate with an
arrow the four groups of codons that conserve the amino acid and with a simple
line those that substitute the amino acid by a new one.

As the code follows evolving it suffers new breaking of symmetry so that the
third (first) base of some codons bring into use or, in the atomic analogy,
some of the fourfold degenerate levels split into two levels\ each one twofold
degenerate. Those levels pointed out with an arrow follow codifying the same
amino acid whereas the other levels substitute it for a new one. Eventually,
in subsequent steps, a few of the twofold degenerate levels split once more
given two non-degenerate levels each. This is the case of codons that codify
methionine (M$_{\text{J}}$), tryptophan (W$_{\text{J}}$) and (again) the stop
signal. The case of isoleucine (I$_{\text{J}}$) is a particular one since the
split level coincides with the twofold one which represents the two codons
that follow codifying the same amino acid. This way, isoleucine is the only
amino acid which is coded by three codons. The stop signal is also threefold
degenerate since it is coded by two groups of codons one twofold degenerate
and the other one non-degenerate. At this step of the evolution the code
frozen to give what would be its present form. It is worth mentioning that the
code evolution gives as a particular result that the amino acids serine
(S$_{\text{J}}$), arginine (R$_{\text{J}}$) and leucine (L$_{\text{J}}$) are
(would be) at present coded by two groups of codons each one. In the three
cases one of the groups is fourfold degenerate and the other one is twofold
degenerate, so that these amino acids are the only three which are sixfold
degenerates. We point out this property in the diagrams with a broken line
linking the two groups of codons. The two groups of codons that codify the
stop signal are also linked by a broken line. We remark again the similarity
of the two evolution diagrams which means that, in our description, the
symmetry breaking and the code freezing must depend of causes other than chirality.

We observe that the diagram of Figure 2 is consistent with the present day
standard genetic code (Figure 4). The diagram of Figure 3, on the other hand,
gives the hypothetical code shown in Figure 5. Actually, this code is not
observed in Nature. We would argue that, although our world (and perhaps the
whole Universe) has the accurate symmetry for the existence of at least two
systems of homochirality, say (\textit{D}-nucleic acids and L-proteins) and
(\textit{L}-nucleic acids and D-proteins), the refinements in the decoding and
synthesis machineries of living organisms, maybe looking for better robustness
and optimization of error-correcting tools, have broken that symmetry by
discriminating between the two systems along the evolution causing the
disappearance of the second one from the Earth.

\begin{figure}[ht!]
 \centering
 \includegraphics[scale=0.50]{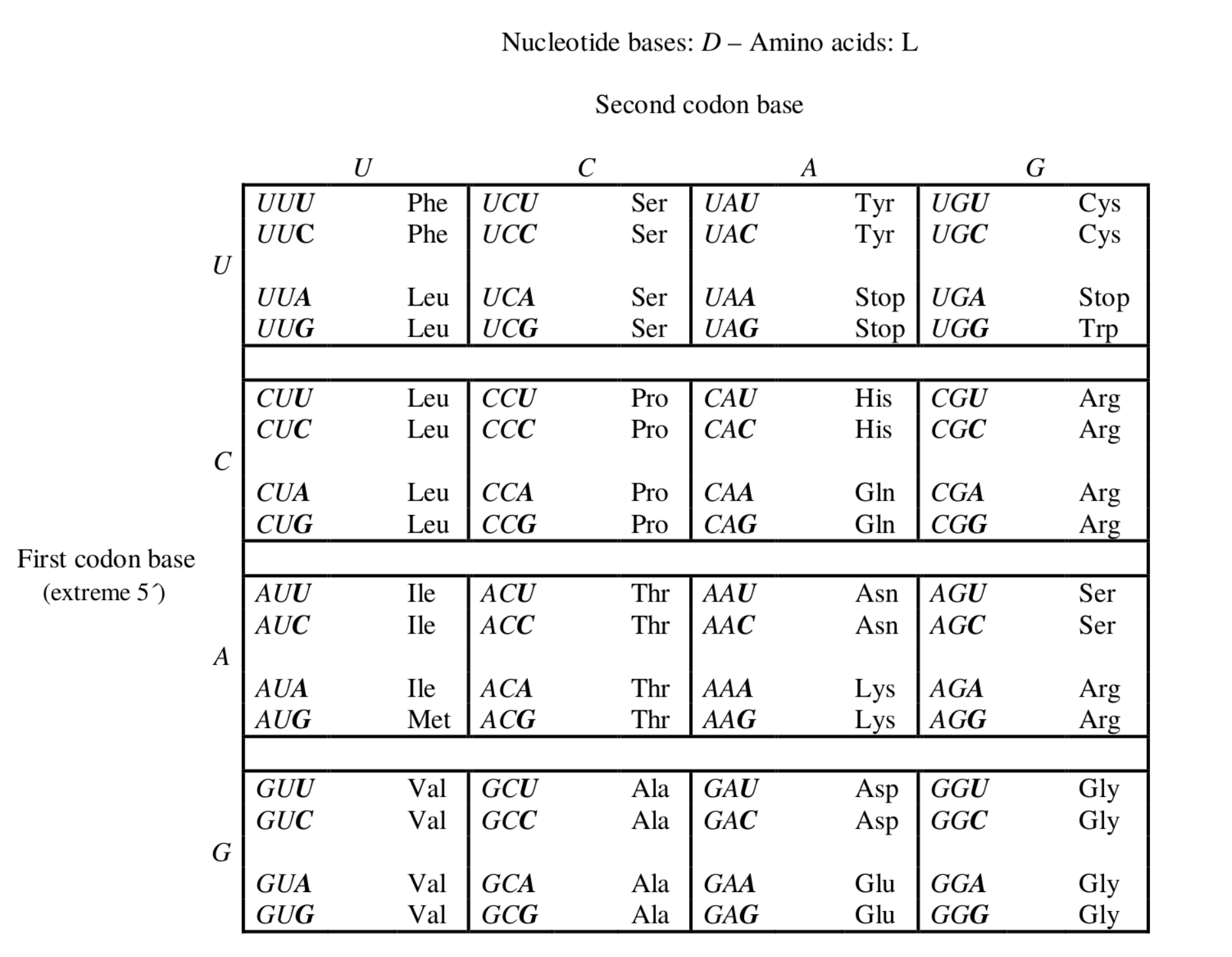}
 \caption{Standard genetic code. The three letters convention for the
amino acids is used and the third base in the codons is remarked in bold. For
simplicity the subindices that identify the class of enantiomer have been
ignored. The reading direction of codons is $5\acute{}\rightarrow 3\acute{}$.}
 \label{fig-04}
\end{figure}

\begin{figure}[ht!]
 \centering
 \includegraphics[scale=0.50]{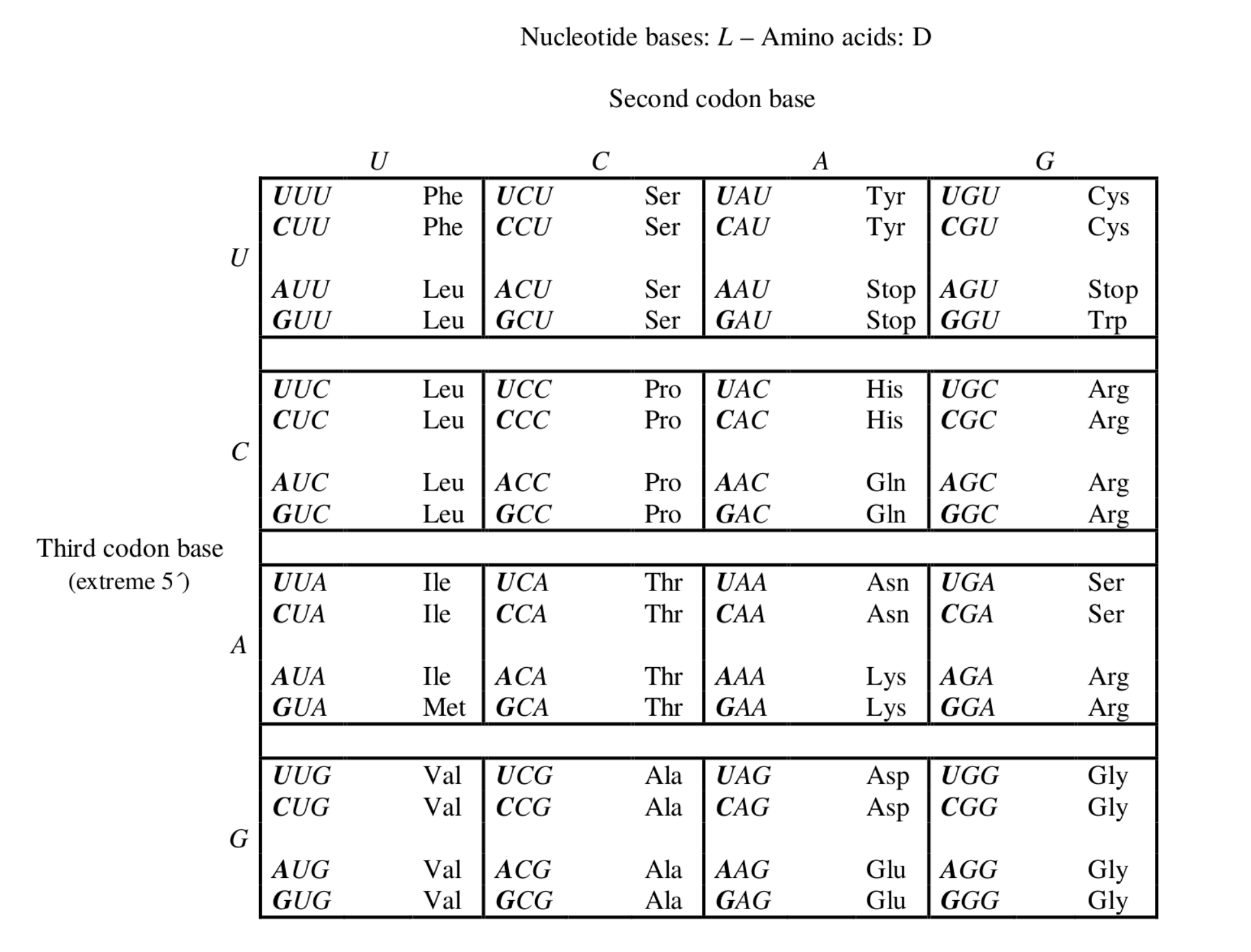}
 \caption{Hypothetical genetic code for the chiral combination
\textit{L}-bases/D-amino acids. The three letters convention for the amino
acids is used and the first base in the codons is remarked in bold. For
simplicity the subindices that identify the class of enantiomer have been
ignored. The reading direction of codons is $3\acute{}\rightarrow 5\acute{}$.}
 \label{fig-05}
\end{figure}

\section{Quaternionic representation of the genetic code and chirality}

Based on the diagrams of Figures 2 and 3, we propose, within a common
formalism, quaternionic representations of the genetic codes corresponding to
both homochiral combinations (\textit{D},L) and (\textit{L},D) according with
the scheme%

\begin{equation}%
\begin{array}
[c]{ccc}%
\mathcal{B}_{\text{\textit{I}}}^{3} & \longrightarrow & \mathcal{A}_{\text{J}%
}\\
\downarrow &  & \downarrow\\
\mathbf{H}_{7,\text{\textit{I}}}^{3}\left(
\mathbb{Z}
\right)  & \longrightarrow & \mathbf{H}\left(
\mathbb{Z}
\right)
\end{array}
\text{ \ \ \ \ \ with \textit{I},J = \textit{D},L and \textit{L},D}
\tag{1}\label{1}%
\end{equation}
where $\mathbf{H}\left(
\mathbb{Z}
\right)  $ denotes the set of integer quaternions (Lipschitz integers),%

\begin{equation}
\mathcal{B}_{\text{\textit{I}}}=\left\{  C_{\text{\textit{I}}}%
,G_{\text{\textit{I}}},U_{\text{\textit{I}}},A_{\text{\textit{I}}}\right\}
\text{ \ \ \ (\textit{I = D,L}),} \tag{2}\label{2}%
\end{equation}

\begin{equation}
\mathcal{A}_{\text{J}}=\left\{  \text{P}_{\text{J}}\text{,A}_{\text{J}%
}\text{,S}_{\text{J}}\text{,T}_{\text{J}}\text{,R}_{\text{J}}\text{,G,C}%
_{\text{J}}\text{,W}_{\text{J}}\text{,L}_{\text{J}}\text{,V}_{\text{J}%
}\text{,F}_{\text{J}}\text{,I}_{\text{J}}\text{,M}_{\text{J}}\text{,H}%
_{\text{J}}\text{,Q}_{\text{J}}\text{,D}_{\text{J}}\text{,E}_{\text{J}%
}\text{,Y}_{\text{J}}\text{,N}_{\text{J}}\text{,K}_{\text{J}}\text{,Stop}%
_{\text{J}}\right\}  \text{ \ \ \ (J\ =\ L,D),\ \ \ } \tag{3}\label{3}%
\end{equation}

\begin{equation}
\mathbf{H}_{7,\text{ \textit{D}}}\left(
\mathbb{Z}
\right)  =\left\{  \left(  2,1,1,1\right)  ,\text{ }\left(  2,-1,1,1\right)
,\text{ }\left(  2,1,-1,1\right)  ,\text{ }\left(  2,1,1,-1\right)  \right\}
\text{ \ \ \ \ \ \ \ \ \ \ \ } \tag{4}\label{4}%
\end{equation}
and%

\begin{equation}
\mathbf{H}_{7,\text{ \textit{L}}}\left(
\mathbb{Z}
\right)  =\left\{  \left(  2,-1,-1,-1\right)  ,\text{ }\left(
2,1,-1,-1\right)  ,\text{ }\left(  2,-1,1,-1\right)  ,\text{ }\left(
2,-1,-1,1\right)  \right\}  . \tag{5}\label{5}%
\end{equation}
Here $\mathcal{B}_{\text{\textit{I}}}^{3}$ is the set of the $64$
\textit{I}-codons (\textit{I = D,L})\ and we assume that the correspondence
$\mathcal{B}_{\text{\textit{I}}}^{3}\rightarrow\mathcal{A}_{\text{J}}$
(\textit{I},J = \textit{D},L; \textit{L},D) are the genetic codes as described
by Figures 4 and 5, whereas the functions $\mathcal{B}_{\text{\textit{I}}}%
^{3}\rightarrow\mathbf{H}_{7,\text{ \textit{I}}}^{3}\left(
\mathbb{Z}
\right)  $ assigns to each \textit{I}-codon a triplet of quaternions of the
sets $\mathbf{H}_{7,\text{ \textit{I}.}}\left(
\mathbb{Z}
\right)  $ (\textit{I = D,L}). Since the norm of all the quaternions of
$\mathbf{H}_{7}\left(
\mathbb{Z}
\right)  $ is $7$, which is a prime number, all the elements of $\mathbf{H}%
_{7}\left(
\mathbb{Z}
\right)  $ are prime quaternions\cite{Davidoff1}.\ Assigning prime quaternions
to the nucleotide bases gives they a certain character of \textit{elemental}
molecules in the present context.

In what follows, in order to simplify the notation, we assign natural numbers
to identify the bases and the amino acids: $C\rightarrow1$, $G\rightarrow2$,
$U\rightarrow3$, $A\rightarrow4$ and P$\rightarrow1$, A$\rightarrow2$,
S$\rightarrow3$, T$\rightarrow4$, R$\rightarrow5$, G$\rightarrow6$,
C$\rightarrow7$, W$\rightarrow8$, L$\rightarrow9$, V$\rightarrow10$,
F$\rightarrow11$, I$\rightarrow12$, M$\rightarrow13$, H$\rightarrow14$,
Q$\rightarrow15$, D$\rightarrow16$, E$\rightarrow17$, Y$\rightarrow18$,
N$\rightarrow19$, K$\rightarrow20$, Stop$\rightarrow21$.

Inspired by the diagrams of Figures 2 and 3 we define the quaternionic functions%

\begin{equation}
\ \ \ \ \
\begin{array}
[c]{c}%
F_{\text{\textit{I}J}}:\mathbf{H}_{7,\text{ \textit{I}.}}^{3}\left(
\mathbb{Z}
\right)  \rightarrow\mathbf{H}\left(
\mathbb{Z}
\right)  \text{ \ \ \ \ \ \ \ \ \ \ \ \ \ \ \ \ \ \ }\\
\text{ \ \ }\left(  q_{\beta\text{\textit{I}}},q_{\gamma\text{\textit{I}}%
},q_{\delta\text{\textit{I}}}\right)  \rightarrow\alpha_{i\text{J}%
}=F_{\text{\textit{I}J}}\left[  \left(  q_{\beta\text{\textit{I}}}%
,q_{\gamma\text{\textit{I}}},q_{\delta\text{\textit{I}}}\right)  \right]
\end{array}
\text{ \ \ (\textit{I},J = \textit{D},L and \textit{L},D)} \tag{6}\label{6}%
\end{equation}
by Eqs. (\ref{7}) and (\ref{8}):%

\begin{align}%
\begin{split}
\text{P}_{\text{L}}\rightarrow\alpha_{1\text{L}}&=q_{1\text{\textit{D}}}q_{1\text{\textit{D}}}\qquad \qquad \qquad \qquad \qquad \text{(}\beta=1\text{, }\gamma=1\text{, }\delta=1\text{,}2\text{,}3\text{,}4\text{)} \\
\text{A}_{\text{L}}\rightarrow\alpha_{2\text{L}}&=q_{2\text{\textit{D}}}q_{1\text{\textit{D}}}\qquad \qquad \qquad \qquad \qquad \text{(}\beta=2\text{, }\gamma=1\text{, }\delta=1\text{,}2\text{,}3\text{,}4\text{)} \\
\text{S}_{\text{L}}\rightarrow\alpha_{3\text{L}}&=q_{3\text{\textit{D}}}q_{1\text{\textit{D}}}=q_{4\text{\textit{D}}}q_{2\text{\textit{D}}}+\gamma_{2\text{\textit{D}};13}\quad\;\;  \text{(}\beta=3\text{, }\gamma=1\text{,}\delta=1\text{,}2\text{,}3\text{,}4\text{ or }\beta=4\text{, }\gamma=2\text{,
}\delta=1\text{,}3\text{)} \\
\text{T}_{\text{L}}\rightarrow\alpha_{4\text{L}}&=q_{4\text{\textit{D}}}q_{1\text{\textit{D}}}\qquad \qquad \qquad \qquad \qquad \text{(}\beta=4\text{, }\gamma=1\text{, }\delta=1\text{,}2\text{,}3\text{,}4\text{)} \\
\text{R}_{\text{L}}\rightarrow\alpha_{5\text{L}}&=q_{1\text{\textit{D}}}q_{2\text{\textit{D}}}=q_{4\text{\textit{D}}}q_{2\text{\textit{D}}}+\gamma_{2\text{\textit{D}};24}\quad\;\; \text{(}\beta=1\text{, }\gamma=2\text{,}\delta=1\text{,}2\text{,}3\text{,}4\text{ or }\beta=4\text{, }\gamma=2\text{,}\delta=2\text{,}4\text{)} \\
\text{G}\rightarrow\alpha_{6}&=\frac{7}{2}\left(q_{2\text{\textit{D}}}q_{2\text{\textit{D}}}+\tilde{q}_{2\text{\textit{D}}}\tilde{q}_{2\text{\textit{D}}}\right) \qquad \quad \;\; \text{(}\beta=2\text{, }\gamma=2\text{,}\delta=1\text{,}2\text{,}3\text{,}4\text{)} \\
\text{C}_{\text{L}}\rightarrow\alpha_{7\text{L}}&=q_{3\text{\textit{D}}}q_{2\text{\textit{D}}}+\gamma_{2\text{\textit{D}};13}\qquad \qquad \qquad \text{(}\beta=3\text{, }\gamma=2\text{, }\delta=1\text{,}3\text{)} \\
\text{W}_{\text{L}}\rightarrow\alpha_{8\text{L}}&=q_{3\text{\textit{D}}}q_{2\text{\textit{D}}}+\gamma_{2\text{\textit{D}};24}+\delta_{2\text{\textit{D}};2}\qquad \;\; \text{(}\beta=3\text{, }\gamma=2\text{, }\delta=2\text{)} \\
\text{L}_{\text{L}}\rightarrow\alpha_{9\text{L}}&=q_{1\text{\textit{D}}}q_{3\text{\textit{D}}}=q_{3\text{\textit{D}}}q_{3\text{\textit{D}}}+\gamma_{3\text{\textit{D}};24} \quad \;\; \text{(}\beta=1\text{, }\gamma=3\text{, }\delta=1\text{,}2\text{,}3\text{,}4\text{ or }\beta=3\text{, }\gamma=3\text{,}\delta=2\text{,}4\text{)} \\
\text{V}_{\text{L}}\rightarrow\alpha_{10\text{L}}&=q_{2\text{\textit{D}}}q_{3\text{\textit{D}}}\qquad \qquad \qquad \qquad \qquad \text{(}\beta=2\text{, }\gamma=3\text{, }\delta=1\text{,}2\text{,}3\text{,}4\text{)} \\
\text{F}_{\text{L}}\rightarrow\alpha_{11\text{L}}&=q_{3\text{\textit{D}}}q_{3\text{\textit{D}}}+\gamma_{3\text{\textit{D}};13} \qquad \qquad \qquad \text{(}\beta=3\text{, }\gamma=3\text{, }\delta=1\text{,}3\text{)} \\
\text{I}_{\text{L}}\rightarrow\alpha_{12\text{L}}&=q_{4\text{\textit{D}}}q_{3\text{\textit{D}}}+\gamma_{3\text{\textit{D}};13}=q_{4\text{\textit{D}}}q_{3\text{\textit{D}}}+\gamma_{3\text{\textit{D}};24}+\delta_{3\text{\textit{D}};4}\quad  \text{(}\beta=4\text{, }\gamma=3\text{,}\delta=1\text{,}3\text{,}4\text{)} \\
\text{M}_{\text{L}}\rightarrow\alpha_{13\text{L}}&=q_{4\text{\textit{D}}}q_{3\text{\textit{D}}}+\gamma_{3\text{\textit{D}};24}+\delta_{3\text{\textit{D}};2} \qquad \;\; \text{(}\beta=4\text{, }\gamma=3\text{, }\delta=2\text{)} \\
\text{H}_{\text{L}}\rightarrow\alpha_{14\text{L}}&=q_{1\text{\textit{D}}}q_{4\text{\textit{D}}}+\gamma_{4\text{\textit{D}};13}\qquad \qquad \qquad \text{(}\beta=1\text{, }\gamma=4\text{, }\delta=1\text{,}3\text{)} \\
\text{Q}_{\text{L}}\rightarrow\alpha_{15\text{L}}&=q_{1\text{\textit{D}}}q_{4\text{\textit{D}}}+\gamma_{4\text{D};24}\qquad \qquad \qquad \text{(}\beta=1\text{, }\gamma=4\text{, }\delta=2\text{,}4\text{)} \\
\text{D}_{\text{L}}\rightarrow\alpha_{16\text{L}}&=q_{2\text{\textit{D}}}q_{4\text{\textit{D}}}+\gamma_{4\text{\textit{D}};13}\qquad  \qquad \qquad \text{(}\beta=2\text{, }\gamma=4\text{, }\delta=1\text{,}3\text{)} \\
\text{E}_{\text{L}}\rightarrow\alpha_{17\text{L}}&=q_{2\text{\textit{D}}}q_{4\text{\textit{D}}}+\gamma_{4\text{\textit{D}};24}\qquad \qquad \qquad \text{(}\beta=2\text{, }\gamma=4\text{, }\delta=2\text{,}4\text{)} \\
\text{Y}_{\text{L}}\rightarrow\alpha_{18\text{L}}&=q_{3\text{\textit{D}}}q_{4\text{\textit{D}}}+\gamma_{4\text{\textit{D}};13}\qquad \qquad \qquad \text{(}\beta=3\text{, }\gamma=4\text{, }\delta=1\text{,}3\text{)} \\
\text{N}_{\text{L}}\rightarrow\alpha_{19\text{L}}&=q_{4\text{\textit{D}}}q_{4\text{\textit{D}}}+\gamma_{4\text{\textit{D}};13}\qquad \qquad \qquad \text{(}\beta=4\text{, }\gamma=4\text{, }\delta=1\text{,}3\text{)} \\
\text{K}_{\text{L}}\rightarrow\alpha_{20\text{L}}&=q_{4\text{\textit{D}}}q_{4\text{\textit{D}}}+\gamma_{4\text{\textit{D}};24}\qquad \qquad \qquad \text{(}\beta=4\text{, }\gamma=4\text{, }\delta=2\text{,}4\text{)} \\
\text{Stop}_{\text{L}}\rightarrow\alpha_{21\text{L}}&=q_{3\text{\textit{D}}}q_{2\text{\textit{D}}}+\gamma_{2\text{\textit{D}};24}+\delta_{2\text{\textit{D}};4}=q_{3\text{\textit{D}}}q_{4\text{\textit{D}}}+\gamma_{4\text{\textit{D}};24}\quad  \text{(}\beta=3\text{, }\gamma=2\text{,}\delta=4\text{ or }\gamma=4\text{, }\delta=2\text{,}
4\text{)}
\end{split}
\tag{7}\label{7}%
\end{align}

\begin{align}
\begin{split}
\text{P}_{\text{D}}\rightarrow\alpha_{1\text{D}}&=q_{1\text{\textit{L}}}q_{1\text{\textit{L}}}\qquad \qquad \qquad \qquad \qquad\text{(}\beta=1\text{,}2\text{,}3\text{,}4\text{, }\gamma=1\text{, }\delta=1\text{)} \\
\text{A}_{\text{D}}\rightarrow\alpha_{2\text{D}}&=q_{1\text{\textit{L}}}q_{2\text{\textit{L}}}\qquad \qquad \qquad \qquad \qquad\text{(}\beta=1\text{,}2\text{,}3\text{,}4\text{, }\gamma=1\text{, }\delta=2\text{)}\\
\text{S}_{\text{D}}\rightarrow\alpha_{3\text{D}}&=q_{1\text{\textit{L}}}q_{3\text{\textit{L}}}=q_{2\text{\textit{L}}}q_{4\text{\textit{L}}}+\gamma_{2\text{\textit{L}};13}\quad\;\;\,  \text{ (}\beta=1\text{,}2\text{,}3\text{,}4\text{,}\gamma=1\text{, }\delta=3\text{ or }\beta=1\text{,}3\text{,
}\gamma=2\text{, }\delta=4\text{)} \\
\text{T}_{\text{D}}\rightarrow\alpha_{4\text{D}}&=q_{1\text{\textit{L}}}q_{4\text{\textit{L}}}\qquad \qquad \qquad \qquad \qquad\text{(}\beta=1\text{,}2\text{,}3\text{,}4\text{, }\gamma=1\text{, }\delta=4\text{)} \\
\text{R}_{\text{D}}\rightarrow\alpha_{5\text{D}}&=q_{2\text{\textit{L}}}q_{1\text{\textit{L}}}=q_{2\text{\textit{L}}}q_{4\text{\textit{L}}}+\gamma_{2\text{\textit{L}};24}\quad\;\;\;\; \text{(}\beta=1\text{,}2\text{,}3\text{,}4\text{, }\gamma=2\text{, }\delta=1\text{ or }\beta=2\text{,}4\text{,
}\gamma=2\text{, }\delta=4\text{)} \\
\text{G}\rightarrow\alpha_{6}&=\frac{7}{2}\left(q_{2\text{\textit{L}}}q_{2\text{\textit{L}}}+\tilde{q}_{2\text{\textit{L}}}\tilde{q}_{2\text{\textit{L}}}\right)  \qquad \quad \;\;\;\; \text{(}\beta=1\text{,}2\text{,}3\text{,}4\text{, }\gamma=2\text{, }\delta
=2\text{)} \\
\text{C}_{\text{D}}\rightarrow\alpha_{7\text{D}}&=q_{2\text{\textit{L}}}q_{3\text{\textit{L}}}+\gamma_{2\text{\textit{L}};13}\qquad \qquad \qquad \; \text{(}\beta=1\text{,}3\text{, }\gamma=2\text{, }\delta=3\text{)} \\
\text{W}_{\text{D}}\rightarrow\alpha_{8\text{D}}&=q_{2\text{\textit{L}}}q_{3\text{\textit{L}}}+\gamma_{2\text{\textit{L}};24}+\delta_{2\text{\textit{L}};2} \qquad \;\;\;\, \text{(}\beta=2\text{, }\gamma=2\text{, }\delta=3\text{)} \\
\text{L}_{\text{D}}\rightarrow\alpha_{9\text{D}}&=q_{1\text{\textit{L}}}q_{3\text{\textit{L}}}=q_{3\text{\textit{L}}}q_{3\text{\textit{L}}}+\gamma_{3\text{\textit{L}};24} \quad \;\;\;\, \text{(}\beta=1\text{,}2\text{,}3\text{,}4\text{, }\gamma=3\text{, }\delta=1\text{ or }\beta=2\text{,}4\text{, }\gamma=3\text{, }\delta=3\text{)} \\
\text{V}_{\text{D}}\rightarrow\alpha_{10\text{D}}&=q_{3\text{\textit{L}}}q_{2\text{\textit{L}}} \qquad \qquad \qquad \qquad \qquad \text{(}\beta=1\text{,}2\text{,}3\text{,}4\text{, }\gamma=3\text{, }\delta=2\text{)} \\
\text{F}_{\text{D}}\rightarrow\alpha_{11\text{D}}&=q_{3\text{\textit{L}}}q_{3\text{\textit{L}}}+\gamma_{3\text{\textit{L}};13}  \qquad \qquad \qquad \; \text{(}\beta=1\text{,}3\text{, }\gamma=3\text{, }\delta=3\text{)} \\
\text{I}_{\text{D}}\rightarrow\alpha_{12\text{D}}&=q_{3\text{\textit{L}}}q_{4\text{\textit{L}}}+\gamma_{3\text{\textit{L}};13}=q_{3\text{\textit{L}}}q_{4\text{\textit{L}}}+\gamma_{3\text{\textit{L}};24}+\delta_{3\text{\textit{L}};4} \quad \text{(}\beta=1\text{,}3\text{,}4\text{, }\gamma=3\text{, }\delta=4\text{)} \\
\text{M}_{\text{D}}\rightarrow\alpha_{13\text{D}}&=q_{3\text{\textit{L}}}q_{4\text{\textit{L}}}+\gamma_{3\text{\textit{L}};24}+\delta_{3\text{\textit{L}};2} \qquad \quad \text{(}\beta=2\text{, }\gamma=3\text{, }
\delta=4\text{)} \\
\text{H}_{\text{D}}\rightarrow\alpha_{14\text{D}}&=q_{4\text{\textit{L}}}q_{1\text{\textit{L}}}+\gamma_{4\text{\textit{L}};13} \qquad \qquad \qquad \; \text{(}\beta=1\text{,}3\text{, }\gamma=4\text{, }\delta=1\text{)} \\
\text{Q}_{\text{D}}\rightarrow\alpha_{15\text{D}}&=q_{4\text{\textit{L}}}q_{1\text{\textit{L}}}+\gamma_{4\text{\textit{L}};24} \qquad \qquad \qquad \; \text{(}\beta=2\text{,}4\text{, }\gamma=4\text{, }\delta=1\text{)} \\
\text{D}_{\text{D}}\rightarrow\alpha_{16\text{D}}&=q_{4\text{\textit{D}}}q_{2\text{\textit{L}}}+\gamma_{4\text{\textit{L}};13} \qquad \qquad \qquad \; \text{(}\beta=1\text{,}3\text{, }\gamma=4\text{, }\delta=2\text{)} \\
\text{E}_{\text{D}}\rightarrow\alpha_{17\text{D}}&=q_{4\text{\textit{L}}}q_{2\text{\textit{L}}}+\gamma_{4\text{\textit{L}};24} \qquad \qquad \qquad \; \text{(}\beta=2\text{,}4\text{, }\gamma=4\text{, }\delta=2\text{)} \\
\text{Y}_{\text{D}}\rightarrow\alpha_{18\text{D}}&=q_{4\text{\textit{L}}}q_{3\text{\textit{L}}}+\gamma_{4\text{\textit{L}};13} \qquad \qquad \qquad \; \text{(}\beta=1\text{,}3\text{, }\gamma=4\text{, }\delta=3\text{)} \\
\text{N}_{\text{D}}\rightarrow\alpha_{19\text{D}}&=q_{4\text{\textit{L}}}q_{4\text{\textit{L}}}+\gamma_{4\text{\textit{L}};13} \qquad \qquad \qquad \; \text{(}\beta=1\text{,}3\text{, }\gamma=4\text{, }\delta=4\text{)} \\
\text{K}_{\text{D}}\rightarrow\alpha_{20\text{D}}&=q_{4\text{\textit{L}}}q_{4\text{\textit{L}}}+\gamma_{4\text{\textit{L}};24} \qquad \qquad \qquad \; \text{(}\beta=2\text{,}4\text{, }\gamma=4\text{, }\delta=4\text{)} \\
\text{Stop}_{\text{D}}\rightarrow\alpha_{21\text{D}}&=q_{2\text{\textit{L}}}q_{3\text{\textit{L}}}+\gamma_{2\text{\textit{L}};24}+\delta_{2\text{\textit{L}};4}=q_{4\text{\textit{L}}}q_{3\text{\textit{L}}}+\gamma_{4\text{\textit{L}};24} \quad \text{(}\beta=4\text{, }\gamma=2\text{, }\delta=3\text{ or }\beta=2\text{,}4\text{, }\gamma=4\text{, }\delta=3\text{)}
\end{split}
\tag{8}\label{8}%
\end{align}

The importance of working with objects that verify a non commutative algebra
is evident from these functions since otherwise amino acids A$_{\text{J}}$ and
R$_{\text{J}}$, and also S$_{\text{J}}$ and L$_{\text{J}}$, would have
associated the same quaternion. The expression for the amino acid glycine (G)
takes into account that it is not chiral. The factor $7$ has to do with the
fact that: $a)$ the norm of the type quaternions can roughly be taken as a
measure of the information needed to codify the corresponding amino acid in
the sense that the larger the norm the larger the necessary information (see
Ref. \cite{Carlevaro1}); $b)$ G is fourfold degenerates and that the norm for
all the other amino acids which are fourfold degenerate is $49$ (see Eqs.
\ref{7}-\ref{10}).

In Eqs. (\ref{7}) and (\ref{8}), the quaternions $\gamma_{\text{i\textit{D}%
;jk}}$ ($\gamma_{\text{i\textit{L};jk}}$) accounts for the level splitting
when the second base of codon is i and the third (first) base is jk$=$ $13$
($C_{\text{\textit{I}}}U_{\text{\textit{I}}}$) or $24$ ($G_{\text{\textit{I}}%
}A_{\text{\textit{I}}}$) \textit{I = D }\ (\textit{ L}). Analogously, the
quaternion $\delta_{\text{i\textit{D}:j}}$ ($\delta_{\text{i\textit{L}:j}}$)
accounts for the level splitting when the second base of the codon is i and
the third (first) base is j$=2$ ($G_{\text{\textit{I}}}$) or $4$%
\ ($A_{\text{\textit{I}}}$) \textit{I = D }\ (\textit{ L}). Thus, in principle
we have as unknown quaternions $\gamma_{2\text{\textit{I}};13}$,
$\gamma_{2\text{\textit{I}};24}$, $\gamma_{3\text{\textit{I}};13}$,
$\gamma_{3\text{\textit{I}};24}$, $\gamma_{4\text{\textit{I}};13}$,
$\gamma_{4\text{\textit{I}};24}$ and $\delta_{2\text{\textit{I}};2}$,
$\delta_{2\text{\textit{I}};4}$, $\delta_{3\text{\textit{I}};2}$ and
$\delta_{3\text{\textit{I}};4}$. Of these $10$ \ unknown quaternions we can
find $5$, say $\gamma_{2\text{\textit{I}};13}$, $\gamma_{2\text{\textit{I}%
};24}$, $\gamma_{3\text{\textit{I}};13}$, $\gamma_{3\text{\textit{I}};24}$,
$\gamma_{4\text{\textit{I}};24}$, by requiring that those amino acids which
are coded by two different groups of codons (case of codons sixfold
degenerates or codons that codify the stop signal) have associated an unique
quaternion and also that the two ways to reach isoleucine (I)\ give the same
quaternion (see Figure 2 (Figure3)). To obtain the quaternions $\delta
_{2\text{\textit{I}};2}$, $\delta_{2\text{\textit{I}};4}$, $\delta_{3;2}$ and
$\delta_{3\text{\textit{I}};4}$ we have assigned to those levels that can not
split more (non degenerate levels) the product of the quaternions associated
with each of the corresponding bases: $\alpha_{8\text{J}}=q_{3\text{\textit{I}%
}}q_{2\text{\textit{I}}}q_{2\text{\textit{I}}}$; $\alpha_{13\text{J}%
}=q_{4\text{\textit{I}}}q_{3\text{\textit{I}}}q_{2}$; $\alpha_{21\text{J}%
}=q_{3\text{\textit{I}}}q_{2\text{\textit{I}}}q_{4\text{\textit{I}}}$;
$\alpha_{12\text{J}}=q_{4\text{\textit{I}}}q_{3\text{\textit{I}}%
}q_{4\text{\textit{I}}}$. Finally for the remaining unknown quaternion
$\gamma_{4\text{\textit{I}};13}$ we have proposed $\gamma_{4\text{\textit{I}%
};13}=-\gamma_{4\text{\textit{I}};24}$.

Taking: $q_{1\text{\textit{D}}}=\left(  2,1,1,1\right)  $,
$q_{2\text{\textit{D}}}=\left(  2,-1,1,1\right)  $, $q_{3\text{\textit{D}}%
}=\left(  2,1,-1,1\right)  $ and $q_{4\text{\textit{D}}}=\left(
2,1,1,-1\right)  $ in Eq.(\ref{7}), we have explicitly obtained%

\begin{equation}%
\begin{array}
[c]{lll}%
\alpha_{1\text{L}}=\left(1,4,4,4\right) & \alpha_{8\text{L}}=\left(  6,-15,-1,9\right) & \alpha_{15\text{L}}=\left(
16,-3,7,1\right) \\
\alpha_{2\text{L}}=\left(3,0,6,2\right) & \alpha_{9\text{L}}=\left(3,6,0,2\right) & \alpha_{16\text{L}}=\left(  -8,3,3,-3\right) \\
\alpha_{3\text{L}}=\left(3,2,0,6\right) & \alpha_{10\text{L}}=\left(5,2,2,4\right) & \alpha_{17\text{L}}=\left(18,-7,5,-1\right) \\
\alpha_{4\text{L}}=\left(3,6,2,0\right) & \alpha_{11\text{L}}=\left(  2,17,1,3\right) & \alpha_{18\text{L}}=\left(-8,9,1,1\right) \\
\alpha_{5\text{L}}=\left(3,0,2,6\right) & \alpha_{12\text{L}}=\left(6,17,3,-3\right) & \alpha_{19\text{L}}=\left(-12,9,3,-5\right) \\
\alpha_{6}=\left(7,0,0,0\right) & \alpha_{13\text{L}}=\left(18,3,-1,3\right) & \alpha_{20\text{L}}=\left(14,-1,5,-3\right) \\
\alpha_{7\text{L}}=\left(3,-2,-6,8\right) & \alpha_{14\text{L}}=\left(-10,7,5,-1\right) & \alpha_{21\text{L}}=\left(18,-1,3,3\right).
\end{array}
\tag{9}\label{9}%
\end{equation}
Analogously, taking $q_{1\text{\textit{L}}}=\tilde{q}_{1\text{\textit{D}}%
}=\left(  2,-1,-1,-1\right)  $, $q_{2\text{\textit{L}}}=\tilde{q}%
_{2\text{\textit{D}}}=\left(  2,1,-1,-1\right)  $, $q_{3\text{\textit{L}}%
}=\tilde{q}_{3\text{\textit{D}}}=\left(  2,-1,1,-1\right)  $ and
$q_{4\text{\textit{L}}}=\tilde{q}_{4\text{\textit{D}}}=\left(
2,-1,-1,1\right)  $ in Eq.(\ref{8}) we have%

\begin{equation}%
\begin{array}
[c]{lll}%
\alpha_{1\text{D}}=\left(1,-4,-4,-4\right) & \alpha_{8\text{D}}=\left(6,15,1,-9\right) & \alpha_{15\text{D}}=\left(16,3,-7,-1\right) \\
\alpha_{2\text{D}}=\left(3,0,-6,-2\right) & \alpha_{9\text{D}}=\left(3,-6,0,-2\right)  &
\alpha_{16\text{D}}=\left(-8,-3,-3,3\right) \\
\alpha_{3\text{D}}=\left(3,-2,0,-6\right) & \alpha_{10\text{D}}=\left(5,-2,-2,-4\right) &
\alpha_{17\text{D}}=\left(18,7,-5,1\right) \\
\alpha_{4\text{D}}=\left(3,-6,-2,0\right) & \alpha_{11\text{D}}=\left(2,-17,-1,-3\right) &
\alpha_{18\text{D}}=\left(-8,-9,-1,-1\right) \\
\alpha_{5\text{D}}=\left(3,0,-2,-6\right) & \alpha_{12\text{D}}=\left(6,-17,-3,3\right)  &
\alpha_{19\text{D}}=\left(-12,-9,-3,5\right) \\
\alpha_{6}=\left(7,0,0,0\right)  & \alpha_{13\text{D}}=\left(18,-3,1,-3\right) & \alpha_{20\text{D}}=\left(14,1,-5,3\right) \\
\alpha_{7\text{D}}=\left(3,2,6,-8\right) & \alpha_{14\text{D}}=\left(-10,-7,-5,1\right) &
\alpha_{21\text{D}}=\left(18,1,-3,-3\right).
\end{array}
\tag{10}\label{10}%
\end{equation}
We denote the set of quaternions assigned to the amino acids as given by
Eqs.(\ref{9}) and (\ref{10}) by $\mathbf{H}_{\alpha\text{D}}\left(
\mathbb{Z}
\right)  $ and $\mathbf{H}_{\alpha\text{L}}\left(
\mathbb{Z}
\right)  $, respectively. We see that the elements of $\mathbf{H}%
_{\alpha\text{\textit{D}}}\left(
\mathbb{Z}
\right)  $ and $\mathbf{H}_{\alpha\text{\textit{L}}}\left(
\mathbb{Z}
\right)  $ verify $\alpha_{i\text{D}}$ $=$ $\tilde{\alpha}_{i\text{L}}$
\ ($i=1,2,\cdots,20,21$) say, the quaternions assigned to both enantiomers of
a given amino acid are mutually conjugates.

\section{Folding of L- and D-proteins in the quaternions formalism}

We have presented quaternionic representations of the standard genetic code
for living systems, where the chiral combination (\textit{D-}bases\textit{/}%
L-amino acids) is preferred, and also of a hypothetical genetic code for
systems, in which we assume that the combination (\textit{L-}bases\textit{/}%
D-amino acids) prevails among other possible chiral combinations. These
representations reproduce the structure of the corresponding codes,
particularly their degeneration. However, the fact that distinguishes the
quaternionic representation over most of available mathematical
representations is the resulting assignation of quaternions to the amino
acids. Because of the advantages of using quaternions to describe spatial
rotations, the association of amino acids with quaternions opens new horizons
beyond the genetic code representation. In this context, we consider here the
suitability of this association, together with our characterization of
chirality, to take account of the folding of homochiral proteins formed by
exclusively L or D amino acids.

The primary structure of a J-protein (J = L,D) formed by $N$ J-amino acids is
a sequence A$_{\text{J}1}$,A$_{\text{J}2}$,$\ldots$,A$_{\text{J}N}$ with
A$_{\text{J}i}\in\mathcal{A}_{\text{J}}$. Our aim is to obtain from this
sequence the spatial coordinates of each one of the atoms of all the amino
acids that constitute the protein when this one is in the native -or
functional- state (tertiary structure). For L-proteins, we take as such the
one corresponding to the protein in physiological solution whose coordinates
can be obtained, after crystallization, by application of, for example, X-ray
diffraction methods. That is the case of most of the proteins whose
coordinates are stored at the Protein Data Bank\cite{PDB}. For D-proteins,
since in the laboratories have been synthesized very few such
proteins\cite{Mandal1,Liu1}, we take in these cases as tertiary
structure the mirror-image of the corresponding experimental L-proteins structure.

In principle we restrict ourselves to determine the coordinates for just the
alpha-carbon atoms of the proteins chain which is not a severe restriction
since is known that there exist (at least for L-proteins) very efficient
algorithms for going from this trace representation to the full atoms
one\cite{Rotkiewicz1}. We also take into account that, in our quaternionic
representation, the J-amino acids sequence is expressed as a sequence of
quaternions $p_{\text{J}1}$,$p_{\text{J}2}$,$\ldots$,$p_{\text{J}N}$ with
$p_{\text{J}i}\in\mathbf{H}_{\alpha\text{J}}\left(\mathbb{Z}\right)$. Under these conditions we proceed now to present an algorithm to determine the spatial coordinates of the alpha-carbon atoms of the protein.

First we observe that although adjacent alpha-carbon atoms are not covalently
bonded their distance is notably stable and take very similar values for all
the pairs within a given protein and also for those belonging to different
proteins. So in our calculations we assume that all these distances are equal
to a unique value $d_{C_{\alpha}-C_{\alpha}}=3.80$ \AA . Thus we determine on
the unit sphere with center at the origin a point for each of the amino acids
(alpha-carbon atoms) in the protein sequence. To the last one we assign
directly the origin, the preceding one is located at the intersection between
the axis $z$ and the sphere surface (versor $\widehat{e}_{z}$). To each of the
remaining alpha-carbon atoms we assign a point on the sphere surface that
results of rotating the versor $\widehat{e}_{z}$ (north pole) by a quaternion.
For the $i$th alpha-carbon atom in the J-sequence, the quaternion responsible
of the rotation is denoted $\widehat{\beta}_{\text{J}i\text{ }}$%
($i=3,4,\cdots,N$). We then expand the chain of alpha-carbon atoms from their
location on the sphere into the back-bone protein three dimensional
configuration (see figure 6) by means of the following iterative procedure,
where initially the$\ \mathbf{r}_{j}$\'{}s are on the sphere surface:

\ \ \ \ \ \ \ \ \ \ \ \ \ \ \ \ \ \ \ \ \ \ \ \ \ \ \ \ \ \ \ \ \ \ \ \ \ \ \ \ \ \ \ \ \ \ do
$i=1,N-2$

$\ \ \ \ $%
\ $\ \ \ \ \ \ \ \ \ \ \ \ \ \ \ \ \ \ \ \ \ \ \ \ \ \ \ \ \ \ \ \ \ \ \ \ \ \ \ \ \ \ \ \ \ \delta
\mathbf{r}=\mathbf{r}_{i+1}$

\ \ \ \ \ \ \ \ \ \ \ \ \ \ \ \ \ \ \ \ \ \ \ \ \ \ \ \ \ \ \ \ \ \ \ \ \ \ \ \ \ \ \ \ \ \ \ \ \ \ \ \ do
$j=1,i$

$\ \ \ \ \ \ \ \ $%
\ $\ \ \ \ \ \ \ \ \ \ \ \ \ \ \ \ \ \ \ \ \ \ \ \ \ \ \ \ \ \ \ \ \ \ \ \ \ \ \ \ \ \ \ \ \ \ \mathbf{r}%
_{j}=\mathbf{r}_{j}+\delta\mathbf{r}$

\ \ \ \ \ \ \ \ \ \ \ \ \ \ \ \ \ \ \ \ \ \ \ \ \ \ \ \ \ \ \ \ \ \ \ \ \ \ \ \ \ \ \ \ \ \ \ \ \ \ \ \ end
do

\ \ \ \ \ \ \ \ \ \ \ \ \ \ \ \ \ \ \ \ \ \ \ \ \ \ \ \ \ \ \ \ \ \ \ \ \ \ \ \ \ \ \ \ \ \ \ end
do\ 

According with the algorithm, the distance between adjacent alpha-carbon atoms
is the unit so, to establish the correct distance, we must multiply the final
calculated coordinates by $d_{C_{\alpha}-C_{\alpha}}$.
\begin{center}
\begin{figure}[ht!]
 \centering
 \includegraphics[scale=0.5]{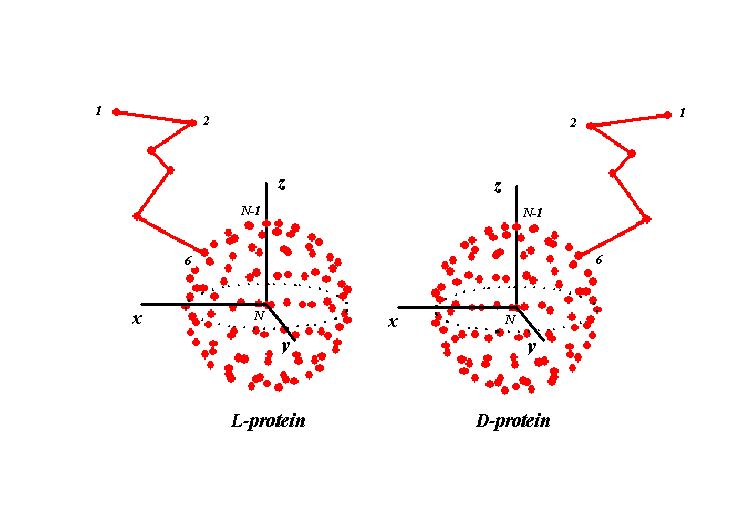}
 \caption{Development of the alpha-carbon atoms backbones of a
hypothetical L-protein of length $N$ and the corresponding D-enantiomer from
their position on the sphere surface into their spatial configuration
(schematic). Both chains are the mirror-image one of the other. In each
spatial chain the last two alpha-carbon atoms, as well as some of the first
ones, are labelled by their order number inside the sequence.}
 \label{fig-06}
\end{figure}

\end{center}

It remains to determine how to calculate the quaternions $\widehat{\beta
}_{\text{J}i\text{ }}$($i=3,4,\cdots,N$). \ In ref. \cite{Carlevaro1} we do
this in a somewhat heuristic way. We take into account that the $i$th amino
acid interacts in some way with the $i-1$ previous amino acids in the sequence
and also with the $N-i$ subsequent ones. Of course that in these interactions
the effect of the medium should be incorporated in some form, for example in
the form of effective interactions between amino acids. Actually we are trying
for a sort of decodification and so we are not directly interested into the
detailed form of the interactions, but we recognize that in any codification
of information that involves those interactions, some trace of their general
form should be. In general it is reasonable to think that the global
interaction includes just two body (effective) interactions so by analogy we
choose with generality for \ $\widehat{\beta}_{\text{J}i\text{ }}$ the
normalized version of the quaternion\ %

\begin{equation}
\beta_{\text{J}i\text{ }}=\sum\limits_{\substack{r=1,2,\cdots,N\\r\neq
i}}c_{\text{J}r}\left(  p_{\text{J}r}\bullet p_{\text{J}i}\right)  \text{
\ \ \ \ \ \ \ \ \ \ \ \ \ \ }\left(  \text{J = L,D; \ }i=1,2,\cdots,N\right)
\tag{11}\label{11}%
\end{equation}
where $\bullet$ denotes the quaternionic dot product:
\[
p_{\text{J}r}\bullet p_{\text{J}i}=\left(  p_{\text{J}r}\right)  _{0}\left(
p_{\text{J}i}\right)  _{0}+\left(  p_{\text{J}r}\right)  _{1}\left(
p_{\text{J}i}\right)  _{1}+\left(  p_{\text{J}r}\right)  _{2}\left(
p_{\text{J}i}\right)  _{2}+\left(  p_{\text{J}r}\right)  _{3}\left(
p_{\text{J}i}\right)  _{3}%
\]
and $c_{\text{J}r}\in\mathbf{H}\left(
\mathbb{R}
\right)  $ ($r=1,2,\cdots,N$) are unknown real quaternions (order quaternions)
which are determined by means of an optimization technique. As such we use the
particle swarm optimization (PSO) procedure of Kennedy and
Eberhart\cite{Kennedy1} taking as function of fitness the difference between
the coordinates of the alpha-carbon atoms calculated following the previous
procedure and the corresponding experimental ones. For these last we take
those directly read from the PDB (for L-proteins) or their mirror-images (for
D-proteins). We take the rmsd (root-mean-square deviation) as a measure of the
fitness difference, using to that effect Bosco K. Ho%
\'{}%
s implementation of Kabsch algorithm\cite{Kabsch1}.

Actually, it is enough to consider J = L since $p_{\text{D}r}=\tilde
{p}_{\text{L}r}$ so, if we take%

\begin{equation}
c_{\text{D}r}=\mathbf{i}c_{\text{L}r}\widetilde{\mathbf{i}\text{ }}%
\mathbf{=}\text{ }\widetilde{\mathbf{i}}c_{\text{L}r}\mathbf{i,}
\tag{12}\label{12}%
\end{equation}
with $\mathbf{i=}\left(  0,1,0,0\right)  $, then a similar relationship is
verified by $\beta_{\text{D}i\text{ }}$:%

\begin{equation}
\beta_{\text{D}i}=\mathbf{i}\beta_{\text{L}i}\widetilde{\mathbf{i}\text{ }%
}\mathbf{=}\text{ }\widetilde{\mathbf{i}}\beta_{\text{L}i}\mathbf{i=}\text{
}\left(  \left(  \beta_{\text{L}i}\right)  _{0},\left(  \beta_{\text{L}%
i}\right)  _{1},-\left(  \beta_{\text{L}i}\right)  _{2},-\left(
\beta_{\text{L}i}\right)  _{3}\right)  . \tag{13}\label{13}%
\end{equation}
We observe that if we denote with $\mathbf{x}_{\text{L}i\text{ }}=\left(
\left(  \mathbf{x}_{\text{L}i\text{ }}\right)  _{1},\left(  \mathbf{x}%
_{\text{L}i\text{ }}\right)  _{2},\left(  \mathbf{x}_{\text{L}i\text{ }%
}\right)  _{3}\right)  $ the point on the sphere that results of rotating the
versor $\widehat{e}_{z}$ by the quaternion $\widehat{\beta}_{\text{L }i\text{
}}$: $\left(  0,\mathbf{x}_{\text{L}i\text{ }}\right)  =\widehat{\beta
}_{\text{L }i\text{ }}\left(  0,\widehat{e}_{z}\right)  \widetilde
{\widehat{\beta}}_{\text{L }i\text{ }}$, \ then the sphere point that results
of rotating the north pole by $\widehat{\beta}_{\text{D }i\text{ }}$is given by%

\begin{equation}
\left(  0,\mathbf{x}_{\text{D}i\text{ }}\right)  =\widehat{\beta}_{\text{D
}i\text{ }}\left(  0,\widehat{e}_{z}\right)  \widetilde{\widehat{\beta}%
}_{\text{D }i\text{ }}=\mathbf{i}\left(  0,\mathbf{x}_{\text{L}i\text{ }%
}\right)  \mathbf{i}=\left(  0,-\left(  \mathbf{x}_{\text{L}i\text{ }}\right)
_{1},\left(  \mathbf{x}_{\text{L}i\text{ }}\right)  _{2},\left(
\mathbf{x}_{\text{L}i\text{ }}\right)  _{3}\right)  , \tag{14}\label{14}%
\end{equation}
say $\mathbf{x}_{\text{D}i\text{ }}$is the mirror-image of $\mathbf{x}%
_{\text{L}i\text{ }}$with respect to the plane $x=0$ in a cartesian axis
$\left(  x,y,z\right)  $ (see figure 6). When we expand the two chains (L and
D) of alpha-carbon atoms from their location on the sphere by using the
previous iterative procedure they result to be the mirror-image one of the other.

In figures 7 to 9 we show the L and D enantiomers of three small proteins as
obtained by using our procedure: in figure 7 of the hormone glucagon (PDB\ ID:
1GCN - length: $29$ amino acids); in figure 8 of the ion channel inhibitor
osk1 toxin (PDB\ ID: 2CK5 - length: $31$ amino acids) and, in figure 9, of a
type III antifreeze protein (PDB ID: 1HG7 - length: $66$ amino acids). In all
the cases the theoretic curves are compared with the corresponding
experimental ones as obtained directly from the PDB for L-proteins and from
the mirror-image of these for D-proteins. The corresponding rmsd%
\'{}%
s for the L-proteins are: $0.103$ \AA \ for 1GCN; $0.091$ \AA \ for 2CK5 and
$0.163$ \AA \ for 1HG7.

\begin{figure}[ht!]
 \centering
 \includegraphics[scale=0.8]{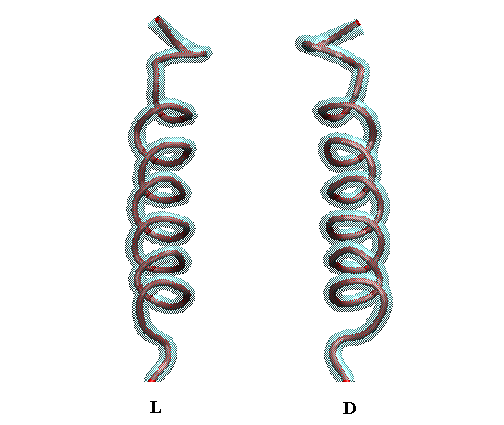}
 \caption{Trace representation of the alpha-carbon atoms backbone for
L- and D-1GCN. Red (dark grey) inner tube: from the coordinates obtained using
our procedure. Cyan (light grey) external transparent tube: from the
coordinates stored at PDB for L and its mirror-image for D.}
 \label{fig-07}
\end{figure}

\begin{figure}[ht!]
 \centering
 \includegraphics[scale=0.75]{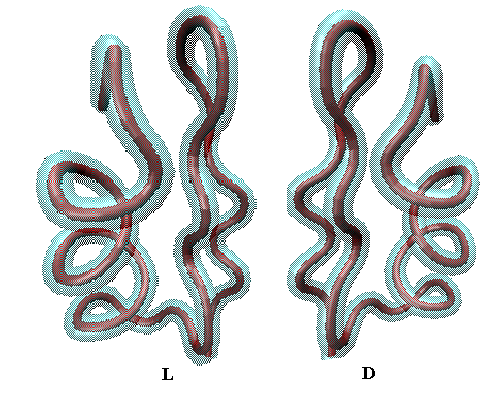}
 \caption{Trace representation of the alpha-carbon atoms backbone for
L- and D-2CK5. Red (dark grey) inner tube: from the coordinates obtained using
our procedure. Cyan (light grey) external transparent tube: from the
coordinates stored at PDB for L and its mirror-image for D.}
 \label{fig-08}
\end{figure}

\begin{figure}[ht!]
 \centering
 \includegraphics[scale=0.9]{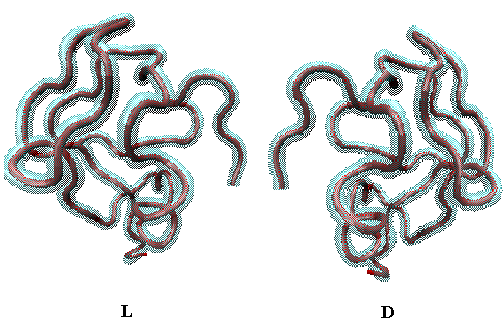}
 \caption{Trace representation of the alpha-carbon atoms backbone for
L- and D-1HG7. Red (dark grey) inner tube: from the coordinates obtained using
our procedure. Cyan (light grey) external transparent tube: from the
coordinates stored at PDB for L and its mirror-image for D.}
 \label{fig-09}
\end{figure}

\section{Remarks}

Assuming that at the beginnings amino acids and nucleotide bases were
synthesized from primordial elements in racemic mixtures and that
\textit{D}-bases were always more affine for L-amino acids whereas
\textit{L}-bases have preferred D-amino acids, we have proposed diagrams for
the evolution of genetic codes which at present would settle the
correspondence between codons and amino acids looking for those affinities.
Actually, of both codes only that corresponding to the affinity system
(\textit{D}-bases/L-amino acids) is nowadays observed on the Earth. Although
the existence of the chiral combination (\textit{L}-bases/D-amino acids) is in
principle possible, none of the organisms that live in our planet shows
\textit{L}-nucleic acids or D-proteins. However, the existence of life in
other planets, including the possibility that it be governed by such a genetic
code (Figure 5), is an open issue.

Our evolution diagrams (Figures 2 and 3) are based on pioneering ideas by
Crick and introduce in a very simple way the concept of broken symmetry. The
fact that the diagrams describe the degeneration breaking and the code
freezing following a similar pattern for both affinities systems,
(\textit{D}-bases/L-amino acids) and (\textit{L}-bases/D-amino acids), means
that we are considering that these aspects basically do not depend of the
molecules chirality but of other physical, chemical and/or biological causes.

Inspired by the evolution diagrams we propose a quaternions based mathematical
representation of the corresponding genetic codes. We assign to each
nucleotide base an integer quaternion, so the codons are triplets of such
quaternions. The representation assigns to each triplet another integer
quaternion that is associated with one of the 20 amino acids (type
quaternions). The bases quaternions belong to the set of eighth prime integer
quaternions of norm $7$ and the nucleotide bases chirality is\ introduced by
partitioning this set into two subsets (Eqs. \ref{4} and \ref{5}) of
cardinality $4$ each with their elements mutually conjugates and associating
their elements with the \textit{D}- and\textit{ L}-bases, respectively. The
correspondences between triplets of quaternions and type quaternions for both
chiral combinations, (\textit{D}/L) and (\textit{L}/D), are given by functions
(Eqs. \ref{7} and \ref{8}) that use the sum and ordinary product of
quaternions and is such that the type quaternions assigned to both enantiomers
of a given amino acid are mutually conjugates.

Apart of preserving the degeneration of the genetic codes as it should be, our
representations distinguish among other mathematical representations of the
genetic code because they assign quaternions to the amino acids as a final
result so that, in view of the close relationship between quaternions and
spatial rotations, a door towards the study of the proteins folding opens. In
this context we propose an algorithm to go from the primary to the tertiary
structure of L- as well as D-proteins. The algorithm uses, besides the integer
type quaternions, a set of real quaternions associated with the order of the
amino acids in the protein sequence. These order quaternions are basically the
same ones for L- and D-proteins so the algorithm is such that for a given
primary sequence the 3D structure of the L- and D-proteins are the
mirror-image one of the other.

Finally, another observation about this algorithm whose critical step is the
building of the quaternion $\beta_{\text{J}i\text{ }}$(Eq. \ref{11}). In ref.
\cite{Carlevaro1} we use for it an expression that involves the ordinary
product between the type quaternion corresponding to the position $i$ and all
the others. However such expression is not adequate for describing with a
common set of order quaternions the folding of L- and D-proteins. To overcome
this problem, in this article, we have changed the ordinary product by a dot
product between type quaternions. As a consequence the number of sets of order
quaternions that adjust a given protein diminishes so that the search for the
unique set that describe the folding of all the proteins (if it exists!) would
be facilitated. We are currently working on this issue.

\textbf{Acknowledgments}

Support of this work by Universidad Nacional de La Plata, Universidad Nacional
de Rosario and Consejo Nacional de Investigaciones Cient\'{\i}ficas y
T\'{e}cnicas of Argentina is greatly appreciated. The authors are members of
CONICET.


\newpage

\begin{thebibliography}{99}                                                                                               %


\bibitem {Palyi1}G. P\'{a}lyi, C. Zucchi and L. Cagliotti (Editors), Advances
in BioChirality (Elsevier Science Ltd., Oxford,1999)

\bibitem {Palyi2}G. P\'{a}lyi, C. Zucchi and L. Cagliotti (Editors), Progress
in Biological chirality (Elsevier Science Ltd., Oxford, 2004)

\bibitem {Bonner1}W.A. Bonner, Homochirality and life, \textit{EXS }85,
159-188 (1998).

\bibitem {Jorissen1}A. Jorissen andC. Cerf, Asymmetric photoreactions as the
origin of biomolecular homochirality:\ A critical review. Origins Life Evol.
Biosphere 32, 129-142 (2002)

\bibitem {Goodman1}G. Goodman and M.E. Gershwin, The origin of life and the
left-handed amino-acids excess: The furthest heavens and the deepest seas?
Exp. Biol. Med. 231, 1587-1592 (2006).

\bibitem {Root-Bernstein1}R.S. Root-Berstein, Simultaneous origin of
homochirality, the genetic code and its direction. Bioessays 29, 689-698 (2007).

\bibitem {Yarus1}M. Yarus, J.G. Caporaso and R. Knight, Origins of the genetic
code:\ The escaped triplet theory. Annu. Rev. Biochem. 74, 179-198 (2005).

\bibitem {Yarus2}M. Yarus, A specific amino acid binding site composed of RNA.
Science 240, 1751-1758 (1988).

\bibitem {Yarus3}M. Yarus, RNA-ligand chemistry: A testable source for thye
genetic code. RNA 6, 475-484 (2000).

\bibitem {Yarus4}I. Majerfield, D. Puthenvendu and M. Yarus, RNA affinity for
molecular L-histidine; genetic code origins. J. Mol Evol. 61, 226-235 (2005).

\bibitem {Yarus5}I. Majerfield and M. Yarus, A diminute and specific RNA
binding site for L-tryptophan. Nucl. Acid, Res. 33, 5482-5493 (2005).

\bibitem {Yarus6}M. Legiewics and M. Yarus, A more complex isoleucine with a
cognate triplet. J. Biol. Chem. 280, 19815-19822 (2005).

\bibitem {Hobish1}M.K. Hobish, N.S. Wickramasinghe and C Ponnamperuma, Direct
interaction between amino acidsand nucleotides as a possible basis for the
origin of the genetic code. Adv. Space Res. 13, 365-382 (1995).

\bibitem {Saxinger1}C. Saxinger, C. Ponnamperuma and C. Woese, Evidence for
the interaction of nucleotides with immobilized amino acids and its
significance for the origin of the genetic code. Nat. New \ Biol. 234, 172-174 (1971).

\bibitem {Walker1}G.W. Walker, Nucleotide-binding site data and the origin of
the genetic code. Biosystems 9, 139-150 (1977).

\bibitem {Root-Bernstein2}R.S. Root-Bernstein, Experimental test of L- and D-
amino acid binding to L- and D- codons suggests that homochirality and codon
directionality emerged with the genetic code. Symmetry 2, 1180-1200 (2010).

\bibitem {Profy1}A.T. Profy and D.A. Usher, Stereoselective aminoacylation of
a dinucleotide monophosphate by imidazolides of DL-alanine and
N-(tert-butoxycarbonyl)-DL-alanine. J- Mol. Evol. 20, 147-156 (1984).

\bibitem {Nandi1}S.D. Banik and N. Nandi, Chirality and protein biosynthesis.
Top. Curr. Chem. (2012).

\bibitem {Englander1}M.T. Englander, J.L. Avins, R.C. Fleisher, B. Liu, P.R.
Effraim, J. Wang, K. Schulten, T.S. Leych, R.L. Gonzalez Jr and V.W. Cornish,
The ribosome can discriminate the chirality of amino acids within its
peptidyl-transferase center. PNAS (In press, 2015).

\bibitem {Carlevaro1}C.M. Carlevaro, R.M. Irastorza and F. Vericat,
Quaternionic representation of the genetic code. ByoSystems (in press) arXiv\_1505.04656v2 [q-Bio.OT].

\bibitem {Crick1}F.H.C. Crick, L. Barnett, S. Brenner and R.J. Watts-Tobin,
General nature of the genetic code for proteins, Nature,192, 1227-1232 (1961).

\bibitem {Crick2}F.H.C. Crick, The origin of the genetic code, J. Mol. Biol.,
38.367-379 (1968).

\bibitem {Hornos1}J.E. Hornos and Y.M.M. Hornos, Algebraic model for the
evolution of the genetic code, Phys. Rev. Letts. 71, 4401-4404 (1993).

\bibitem {Maddox1}J. Maddox, The genetic code by numbers, Nature, 367, p. 111 (1994).

\bibitem {Stewart1}I. Stewart, Broken symmetry in the genetic code?, New
Scientist, 1915, p. 16 (1994).

\bibitem {Hamilton1}W.R. Hamilton, On quaternions; or on a new system of
imaginaries in algebra (letter to John T. Graves dated October 17, 1843) (1843).

\bibitem {Hamilton2}W.R. Hamilton, Elements of quaternions, edited by W.E.
Hamilton (Longmans, Green \& Co., London, 1866).

\bibitem {Altmann1}S.L. Altmann, Rotations, quaternions and double groups
(Dover Publications, Inc, Mineola, New York, 2005 ).

\bibitem {Creighton1}T.E. Creighton (Editor), Protein folding (W.H. Freeman
and Company, New York, 1992).

\bibitem {BenNaim1}A. Ben Naim, The protein folding problem and its solutions
(World Scientific Publishing Co., Singapore, 2013).

\bibitem {Davidoff1}G. Davidoff, P. Sarnak and A. Valette, Elementary number
theory, group theory and Ramanujan graphs (Cambridge University Press,
Cambridge, UK, 2003).

\bibitem {Miller1}S.L. Miller, A production of amino acids under possible
primitive earth conditions, Science, 117, 528-529 (1953).

\bibitem {Parker1}E.T. Parker, H.J. Cleaves, J.P. Dworkin, D.P. Glavin, M;-
Callahan, A. Aubrey, A. Lezcano and J.L. Bada, Primordial synthesis of amines
and amino acids in a 1958 Miller H$_{\text{2}}$S-rich spark discharge
experiment. PNAS USA, 108, 5526-5531 (2011).

\bibitem {Trifonov1}E.N. Trifonov, Consensus temporal order of amino acids and
evolution of the triplet code, Gene, 261, 139-151 (2000).

\bibitem {PDB}Protein Data Bank, http://www.rcsb.org/pdb/

\bibitem {Mandal1}K. Mandal, M. Uppalapati, D. Ault-Rich\'{e}, J. Kenney, J.
Lowitz, S. S. Sidhu, and S. B.H. Kent, Chemical synthesis and X-ray structure
of a heterochiral D-protein antagonist plus vascular endothelial growth factor
protein complex by racemic crystallography.Proc Natl Acad Sci USA 109,
14779--14784 (2012).

\bibitem {Liu1}M. Liu, M. Pazgier, C. Li, W. Yuan, C. Li and W. Lu, A
left-handed solution to peptide inhibition of the p53-MDM2 interaction. Angew
Chem Int Ed Engl, 49, 3649--3652 (2010).

\bibitem {Rotkiewicz1}P. Rotkiewicz and J. Skolnick, Fast procedure for
reconstruction of full-atom protein models from reduced representations, J.
Comp. Chem, 29, 1460-1465 (2008).

\bibitem {Kennedy1}J. Kennedy and R. Eberhart, R. Particle Swarm Optimization.
Proceedings of IEEE International Conference on Neural Networks IV. pp.
1942--1948 (1995).

\bibitem {Kabsch1}W. Kabsch, A solution for the best rotation to relate two
sets of vectors, Acta Cryst. A32, 922-923 (1976).

\newpage
\end{thebibliography}
\end{document}